\documentclass[aip,jap,showpacs,showkeys]{revtex4}
\usepackage{graphicx}
\usepackage{dcolumn}
\usepackage{bm}

\begin{document}

\title[Maximum Spectral Luminous Efficacy of White Light]{Maximum Spectral Luminous Efficacy of White Light}

\author{Thomas W. Murphy, Jr.}
\affiliation{Department of Physics, University of California, San Diego,\\
9500 Gilman Drive, La Jolla, CA 92093-0424, USA.}
\email{tmurphy@physics.ucsd.edu}

\date{\today}

\begin{abstract}

As lighting efficiency improves, it is useful to understand the theoretical limits to luminous efficacy for light that we perceive as white.  Independent of the efficiency with which photons are generated, there exists a spectrally-imposed limit to the luminous efficacy of any source of photons.  We find that, depending on the acceptable bandpass and---to a lesser extent---the color temperature of the light, the ideal white light source achieves a spectral luminous efficacy of 250--370~lm/W.   This is consistent with previous calculations, but here we explore the maximum luminous efficacy as a function of photopic sensitivity threshold, color temperature, and color rendering index; deriving peak performance as a function of all three parameters. We also present example experimental spectra from a variety of light sources, quantifying the intrinsic efficacy of their spectral distributions.

\end{abstract}


\maketitle

\section{Introduction}

Lighting technology has evolved rapidly in the past decade, enabling
replacement of inefficient incandescent light bulbs with compact fluorescent
lights (CFLs) and light-emitting diodes (LEDs). The figure of merit
for the energy efficiency of lighting is \emph{luminous efficacy}, measured
in lumens per watt. Inefficiency can be separated into two main forms:
inefficient production of photons from the input power source; and
distribution of photons outside of the visible spectrum. For example,
incandescent lights are marvelously efficient at generating photons
from the input electrical source, but produce the vast majority of
these photons at near-infrared wavelengths where the human eye has
no sensitivity. We can break the total luminous efficacy into spectral
and electrical components, so that $\eta_{\mathrm{L}}=\eta_{\mathrm{S}}\eta_{\mathrm{E}}$.
The spectral part, $\eta_{\mathrm{S}}$ carries units of lm/W, while
the electrical part, $\eta_{\mathrm{E}}$ is unitless, representing
the ratio of luminous power output to electrical power input. This
paper focuses on the spectral distribution aspect of lighting efficiency,
and will not address the efficiency with which photons are produced.

A monochromatic source at 555~nm---the peak of the photopic sensitivity
curve \cite{photopic}---will produce 683~lm/W of light output \cite{katsuyama}.
For other monochromatic wavelengths, the luminous efficacy is reduced
by a factor according to the photopic function. For instance, at the
633~nm wavelength of helium-neon lasers, the sensitivity of the eye
is only 23.5\% compared to its peak value (leading to 160
lm/W), whereas the human eye has 88.5\% sensitivity at the 532~nm
wavelength of frequency-doubled Nd:YAG lasers, corresponding to an efficacy of
604 lm/W.

But monochromatic illumination---as efficient as it might be---is
often deemed to be unacceptable, providing no color differentiation. Even
a composite of several line sources that may appear white-like can
inadequately render some colored items, as can often be the case under
the line-dominated spectrum of fluorescent lights. 

Sporadic instances in the literature \cite{narukawa-new,chakraborty} place
the maximum luminous efficacy for white light in the range of 250--300~lm/W
without reference to supporting material.  Some papers (e.g.,
\cite{narukawa-old}) display the appropriate integral along with a
numerical result, but without detailing the integration limits or the
specifics of the integrand. Uchida \& Taguchi \cite{uchida} estimate the
theoretical luminous efficacy of a white LED with good color rendering to
be $\sim300$~lm/W.  Coltrin et al.  \cite{coltrin} found that it is
theoretically possible to synthesize a light achieving 408~lm/W with a
color rendering index in excess of 90 using four discrete wavelengths.  But
in practice the color rendering becomes unacceptable when the fluxes of the
individual line sources are allowed to vary from the design by a standard
deviation of 10\%, facing the additional hurdle that the requisite
narrow-line sources are not easily obtained. Yun-Li et al. \cite{yun-li}
explored combinations of LED lights to synthesize white light, finding that
luminous efficacies in excess of 300~lm/W and suitable color rendering
could be achieved with three LEDs.  None of these previous estimates are in
error, but nor do they represent a comprehensive study of the limits to
spectral luminous efficacy as a function of the quality characteristics of
the light.

In this paper, we examine the theoretical limits to (spectral) luminous
efficacy for lights that we would perceive as white, to varying degrees.  We
find that the maximum efficacy is in the range of 250--370~lm/W, and that
the color temperature of the light has only a modest impact within the range
of values typically encountered. We explore lighting performance as a
function of both the bandpass associated with a minimum threshold, and as a
function of color temperature.  We also evaluate the maximum luminous
efficacy as a function of color rendering index.  Finally, we look at the
spectral luminous efficacy of a variety of modern light sources. For what
follows, references to luminous efficacy are to be interpreted as spectral
luminous efficacy unless explicitly noted otherwise.

\section{Method of Calculation}

To assess the theoretical limits of spectral luminous efficacy, we
integrate the normalized spectral density function, $B_{\lambda}$, in fractional
luminous power per meter (of wavelength), times the photopic sensitivity
curve \cite{photopic} of the eye, $\bar{y}(\lambda)$, which lies
between 0.0 and 1.0, peaking at $\lambda=555$~nm:
\begin{equation}
\eta_{\mathrm{S}}=683\int_{0}^{\infty}\bar{y}(\lambda)B_{\lambda}d\lambda.\label{eq:eta_s-def}
\end{equation}
 The factor of 683 in front scales the result in accordance with the
definition for the lumen.  The function, $\bar{y}(\lambda)$, peaks at a
value of 1.0 at $\lambda = 555$~nm.

\begin{figure}
\begin{center}\includegraphics[scale=0.75]{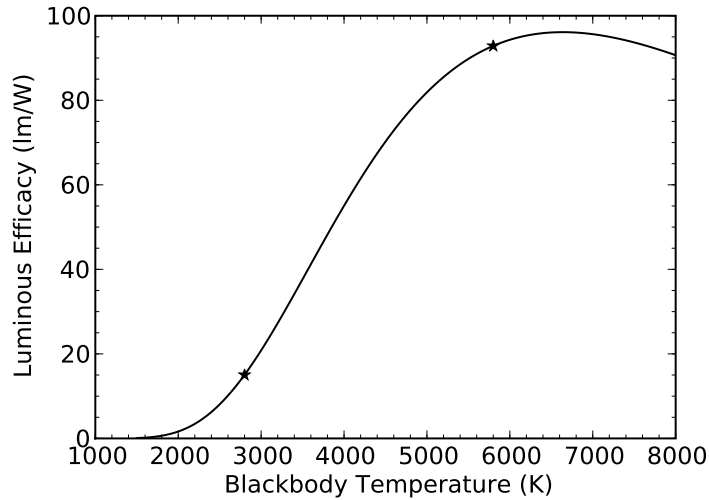}\end{center}

\caption{Luminous efficacy of pure blackbody radiation as a function of temperature.
The effective temperatures of a typical incandescent filament and
the solar surface are marked by stars, corresponding to 15 and 93~lm/W,
respectively. \label{fig:planck}}
\end{figure}

The Sun is our standard ``white'' light source, and is well approximated
by a Planck blackbody at the Sun's surface temperature of 5800~K.
The Planck function---normalized so that the integral over all wavelengths
is unity---is:
\begin{equation}
B_{\lambda}=15\left(\frac{hc}{\pi kT}\right)^{4}\lambda^{-5}\left[ e^{hc/\lambda kT}-1\right] ^{-1}\,\,\mathrm{m^{-1}},\label{eq:planck}
\end{equation}
where $\lambda$ is wavelength, $h$ is Planck's constant, $c$ is
the speed of light, $k$ is Boltzman's constant, and $T$ is the blackbody
temperature, in Kelvin. A blackbody at the temperature of the Sun
results in a luminous efficacy of $\eta_{\mathrm{S}}=93$~lm/W. Only
37\% of its light falls within the visible band from 400~nm to 700~nm.
By comparison, a light bulb at 2800~K has a luminous efficacy---determined
in the manner above---of 15~lm/W, with 6\% of its light in the 400--700~nm
band. Fig.~\ref{fig:planck} shows the luminous efficacy as a function
of blackbody temperature, peaking at 6640~K and 96.1~lm/W. 
As a caveat that does not impact the conclusions of this paper, we note
that actual tungsten filaments achieve 15~lm/W at lower temperatures around
2500~K---being imperfect blackbodies that selectively emit light
in the visible compared to the infrared part of the spectrum. 

To evaluate the color rendering performance of each source, we compute the
correlated color temperature (CCT), offset from ``white,'' and color
rendering index (CRI) for the spectra considered herein---following the
procedure outlined by the International Commission on Illumination (CIE)
\cite{cie}.  In brief, the procedure involves establishing the chromatic
coordinates of the source within the International Commission on
Illumination (known as the CIE) 1960 $uvY$ color space, and relating this
to the chromatic locus of points corresponding to blackbody radiation
profiles. The CCT is the closest blackbody temperature in color space, and
the Pythagorean distance between the points indicates how ``white'' the
source looks ($>5.4\times10^{-3}$ is considered to be too far for the CRI
computation to be reliable).  Establishing the CRI is substantially more
involved, and is based on comparing a standard set of 8 Munsell colors
under illumination by both the spectrum under study and by a blackbody at
the CCT---ultimately forming the offset metric in the CIE 1964 $UVW$ color
space. The CRI ranges from 0--100, with values above 90 typically
considered to be adequate for general lighting. The CRI as calculated in
this way is a flawed construct, but nonetheless is in common practice and
is used here for comparative purposes.

If we construct a \emph{truncated} 5800~K blackbody so that it emits
light only in the range between cutoff wavelengths
$\lambda_{\mathrm{blue}}$ and $\lambda_{\mathrm{red}}$,
the luminous efficacy is now given by:
\begin{equation}
\eta_{\mathrm{S}}=683\frac{\int_{\lambda_{\mathrm{blue}}}^{\lambda_{\mathrm{red}}}\bar{y}(\lambda)B_{\lambda}d\lambda}{\int_{\lambda_{\mathrm{blue}}}^{\lambda_{\mathrm{red}}}B_{\lambda}d\lambda},\label{eq:trunc-eta_s}
\end{equation}
and evaluates to 251~lm/W for a 5800~K blackbody truncated to emit
light only between $\lambda_{\mathrm{blue}}=400$~nm and $\lambda_{\mathrm{red}}=700$~nm.
This describes one manifestation of the ideal white light, which might
in principle be synthesized out of narrow emission sources, such as
LEDs, over a range of wavelengths. 

\begin{figure}
\begin{center}\includegraphics[scale=0.75]{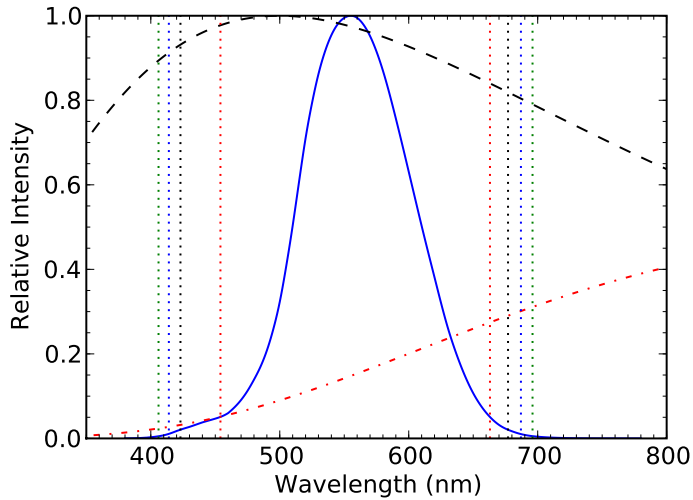}\end{center}

\caption{(Color online) The photopic sensitivity curve is shown as the solid curve.
Blackbodies corresponding to 5800~K (dashed) and 2800~K (dash-dot),
normalized to the same total radiant flux, are shown for reference.  Dotted
vertical lines represent different bounds within which we consider
artificial ideal white light sources, corresponding to photopic curve
intercepts at 0.5\%, 1\%, 2\%, and 5\% sensitivity
levels.\label{fig:curves}}

\end{figure}

More sensible would be to base the $\lambda_{\mathrm{blue}}$ and $\lambda_{\mathrm{red}}$
limits on human photopic sensitivity levels, rather than on the arbitrary---although
convenient---400~nm and 700~nm values. For the analysis in this
paper, we principally concentrate on the part of the photopic response
that lies above 0.5\%, 1\%, 2\%, and 5\% of peak sensitivity. These
threshold levels correspond to $\lambda_{\mathrm{blue}}$ values of
405.7, 413.2, 422.3, and 453.2~nm, respectively, and to $\lambda_{\mathrm{red}}$
values of 697.0, 687.4, 677.4, and 663.2~nm, respectively. Fig.~\ref{fig:curves}
illustrates these ranges in relation to the photopic sensitivity curve,
together with blackbody curves at 2800~K and 5800~K for reference. 

\section{Results}

Using the four principal wavelength ranges discussed above, the ideal
luminous efficacy for light following a 5800~K spectrum computes to $\eta_{\mathrm{S}}=260$,
274, 292, and 348~lm/W, corresponding to 0.5\%, 1\%, 2\%, and 5\%
sensitivity thresholds, respectively. Tab.~\ref{tab:truncated}
lists additional properties of these truncated spectra, together with
two other cases. The results are rather similar for a truncated 2800~K
source. We see in Tab.~\ref{tab:truncated} some justification for
stopping at a 5\% photopic cutoff, as the CRI for these cases abruptly
enters a region considered to be unacceptable in the lighting industry.
Perhaps more importantly, the $u$--$v$ chromaticity offset for these
cases exceeds the threshold of $5.4\times10^{-3}$, and therefore
would no longer be considered ``white.''  It should be noted that the CRI is
not customarily compared among sources with differing color temperatures.

\begin{table*}

\caption{Properties of truncated continuum sources.\label{tab:truncated}}

\begin{ruledtabular}
\begin{tabular}{lccccc}
Source&
photopic cutoff&
CCT (K)&
$\eta_{\mathrm{S}}$ (lm/W)&
Planckian offset&
CRI\\
\hline
5800~K blackbody&
0.5\%&
5784&
260&
$0.64\times 10^{-3}$&
99.4\\
&
1\%&
5756&
274&
$1.3\times 10^{-3}$&
98.8\\
&
2\%&
5653&
292&
$3.4\times 10^{-3}$&
96.5\\
&
5\%&
4646&
348&
$25\times 10^{-3}$&
68.6\\
2800~K blackbody&
0.5\%&
2811&
256&
$0.14\times 10^{-3}$&
99.4\\
&
1\%&
2821&
276&
$0.46\times 10^{-3}$&
98.7\\
&
2\%&
2839&
299&
$1.3\times 10^{-3}$&
97.2\\
&
5\%&
2831&
343&
$10\times 10^{-3}$&
82.7\\
uniform $B_{\lambda}$&
0.5\%&
5440&
253&
$3.7\times 10^{-3}$&
96.4\\
&
1\%&
5415&
268&
$2.9\times 10^{-3}$&
97.1\\
&
2\%&
5324&
287&
$0.71\times 10^{-3}$&
98.4\\
&
5\%&
4418&
344&
$21\times 10^{-3}$&
71.5\\
\end{tabular}
\end{ruledtabular}
\end{table*}

Another case presented in Tab.~\ref{tab:truncated} is that of a truncated
spectrally uniform light source, with constant $B_{\lambda}$ across the
range from $\lambda_{\mathrm{blue}}$ to $\lambda_{\mathrm{red}}$, and no
light outside of this range. For the four principal sensitivities examined
here, the uniform case delivers 253, 268, 287, and 344~lm/W. These numbers
are very similar to the blackbody cases considered above. It is interesting
that the Planckian offset (departure from ``white'') and the CRI initially
improve as the photopic cutoff increases. The uniform spectrum is red- and
blue-heavy compared to a 5800~K blackbody, so that increasing truncation
initially compensates this overabundance of light at the extremes.  All the
same, by the time one reaches a 5\% photopic cutoff, the light is no longer
acceptable, either in terms of Planckian offset or CRI.

\begin{figure}
\begin{center}\includegraphics[scale=0.75]{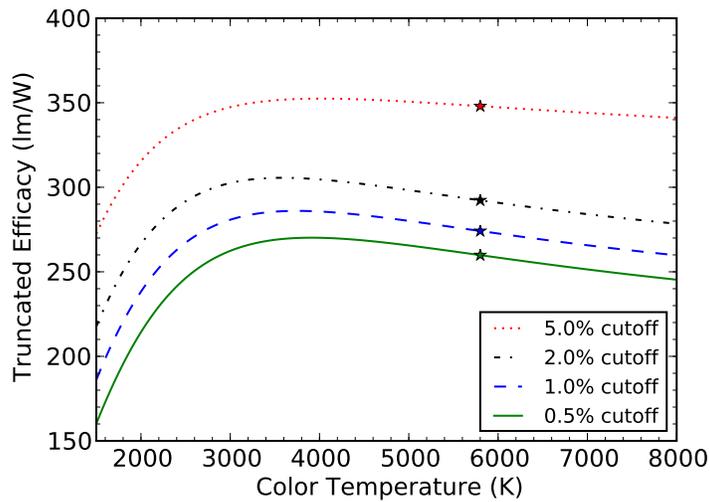}\end{center}

\caption{(Color online) Luminous efficacy of ideal, truncated sources for
four different wavelength ranges as a function of blackbody temperature.
The four wavelength ranges corresponding to 0.5\%, 1\%, 2\%, and 5\%
sensitivity thresholds of the photopic response curve, whose wavelengths
are indicated by vertical dotted lines in Fig.~\ref{fig:curves}. Stars mark
the efficacies at 5800~K.  \label{fig:eff-trunc}}

\end{figure}

We have become accustomed to incandescent lighting, and have developed
some affinity for the ``warm'' color temperatures they emit. A
daylight spectrum presented at night may be perceived as jarring and
harsh. Investigating the maximum luminous efficacy for lights of different
color temperatures reveals a surprise---perhaps partly exposed by
the similarity of numbers in the two blackbody and uniform spectral density
cases above. The maximum luminous efficacy in a truncated spectral
source is largely independent of color temperature, as seen in Fig.~\ref{fig:eff-trunc}.

\begin{figure*}
\includegraphics[%
  scale=0.6]{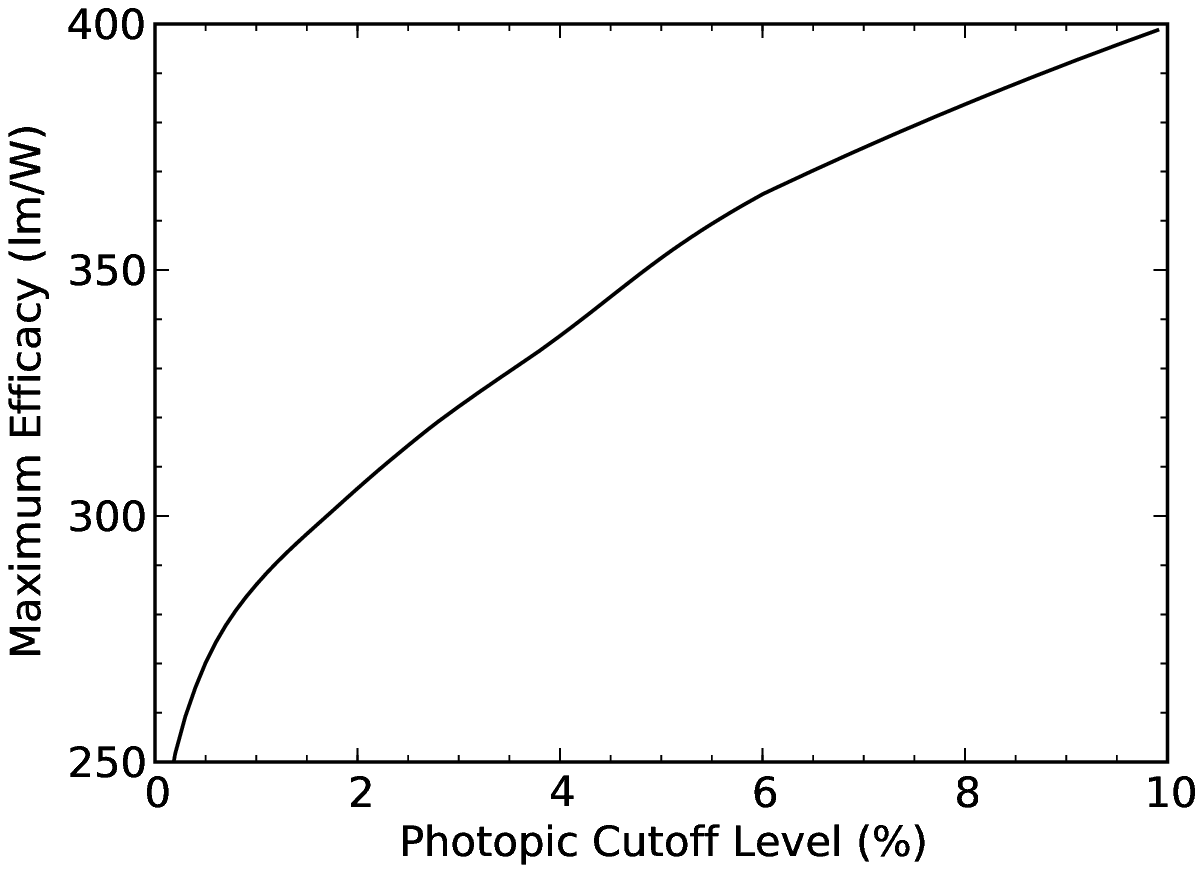}\hfill{}\includegraphics[%
  scale=0.6]{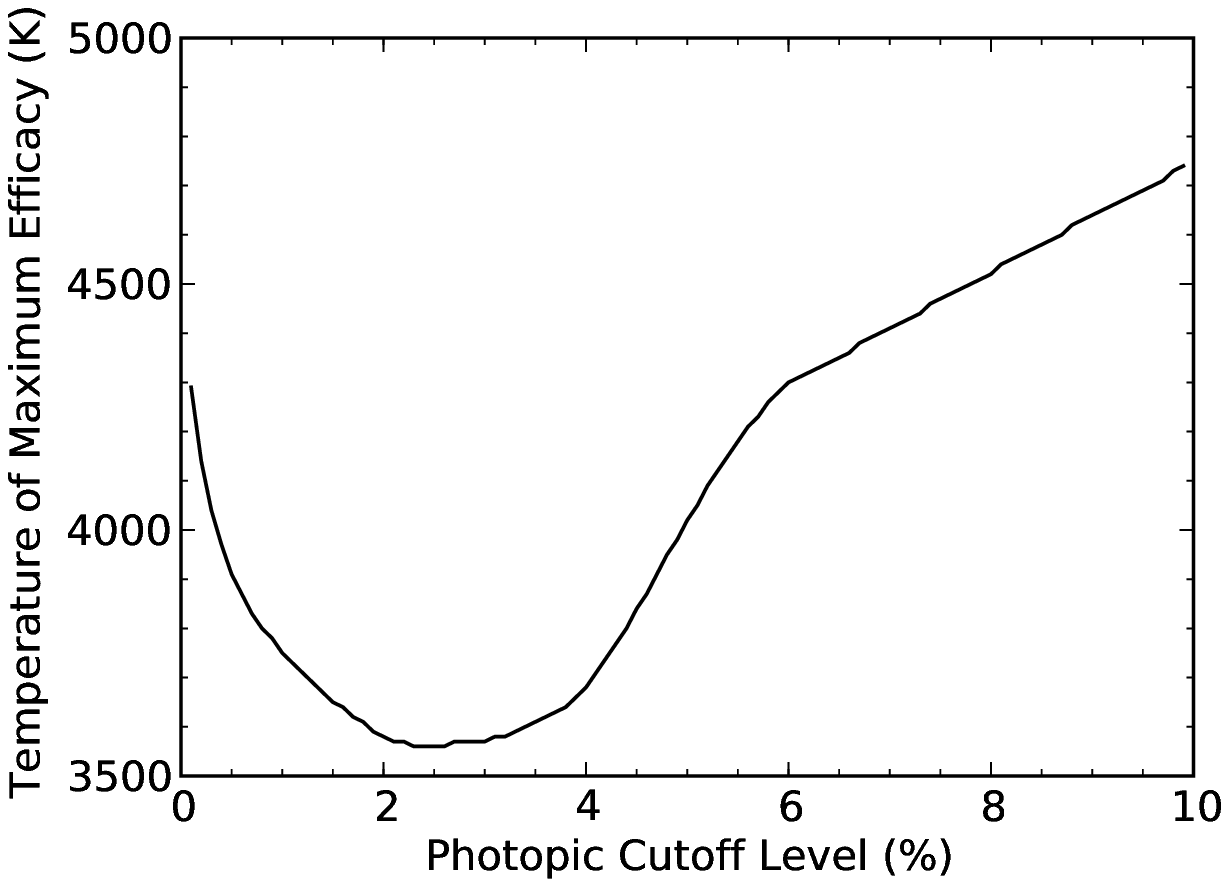}

\caption{Maximum luminous efficacy achievable as a function of photopic
sensitivity threshold (left) and the corresponding color temperature
(right).\label{fig:max_eff}}

\end{figure*}

Between color temperatures of 2500--8000~K, the maximum luminous efficacy
for a particular spectral cutoff varies by less than 10\%.  It is
interesting to ascertain the maximum efficacy achievable as a function of
sensitivity threshold, as well as the temperature at which maximum efficacy
is realized. The result is presented in Fig.~\ref{fig:max_eff}.  The
reversal in peak color temperature as the photopic cutoff increases beyond
about 3\% can be seen as a result of the flattening curves in
Fig.~\ref{fig:eff-trunc}.  The maximum of each curve first shifts toward
lower color temperatures as the photopic cutoff increases, but then the
curves begin to flatten at high color temperatures, pushing the maximum to
higher temperatures.  The break seen at 6\% in both panels of
Fig.~\ref{fig:max_eff} relates to the point at which the blue
tail of the photopic curve abruptly transitions to a higher slope.

\begin{figure*}
\includegraphics[%
  scale=0.6]{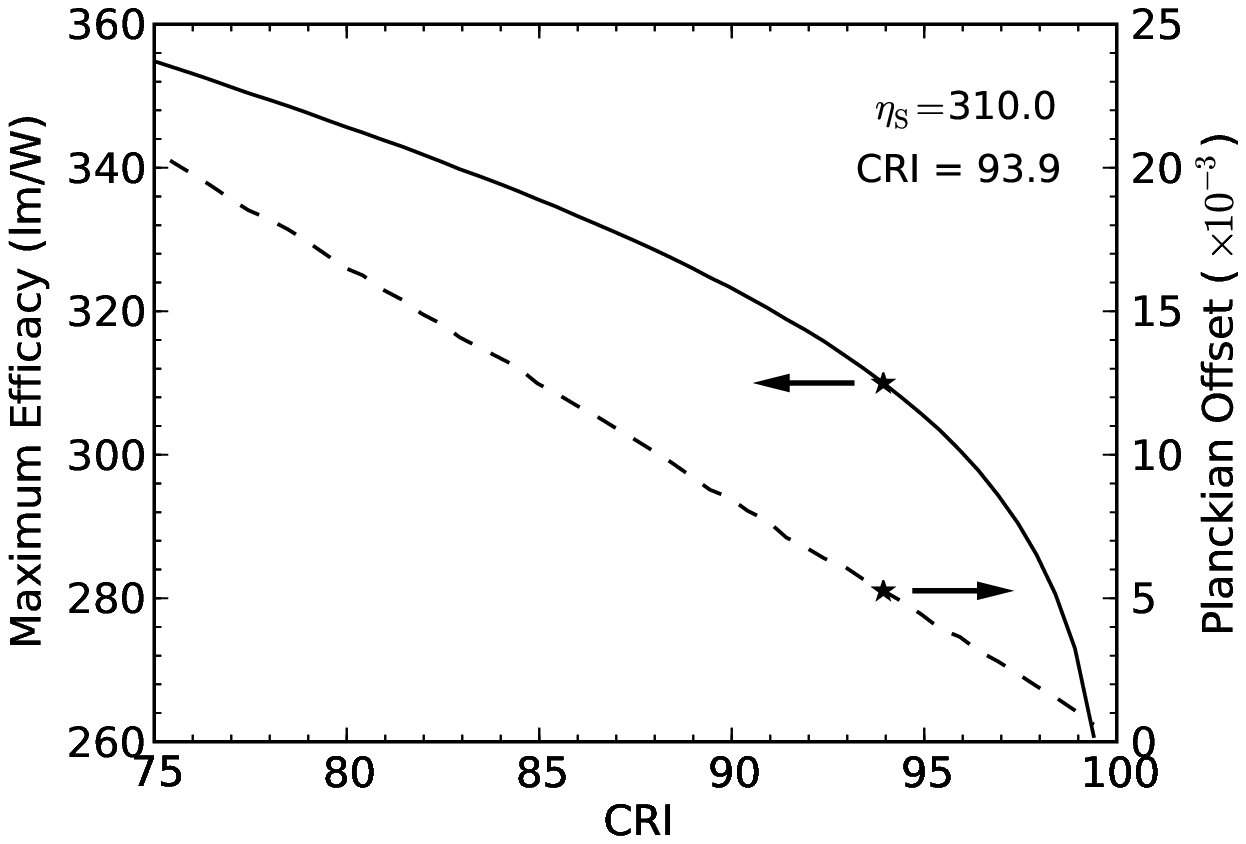}\hfill{}\includegraphics[%
  scale=0.6]{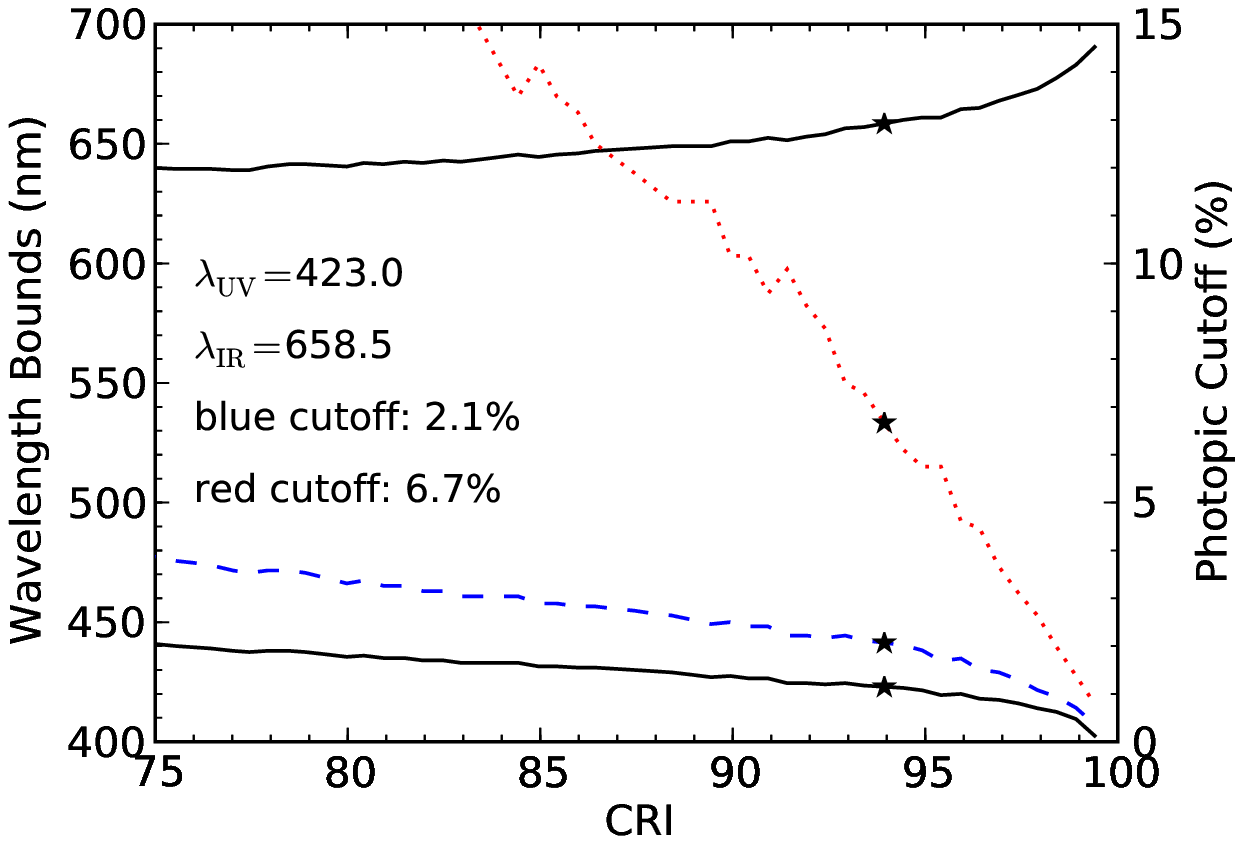}

\caption{(Color online) Conditions for maximum luminous efficacy for a
truncated blackbody at 5800~K, allowing asymmetric red and blue cutoff
wavelengths.  At left is the maximum luminous efficacy (solid) and
associated Planckian offset (dashed) as a function of CRI achieved.  The
star marks the point at which the Planckian offset (central panel) reaches
$5.4\times 10^{-3}$ and is no longer considered to be ``white."  The plot at
right shows the wavelength cutoffs (solid curves) that maximize
$\eta_{\mathrm{S}}$, and their corresponding photopic cutoff sensitivities
(dashed for blue, dotted for red).  Numbers correspond to the stars, which
themselves indicate the point at which the Planckian offset reaches
$5.4\times 10^{-3}$. \label{fig:max58}}

\end{figure*}

\begin{figure*}
\includegraphics[%
  scale=0.6]{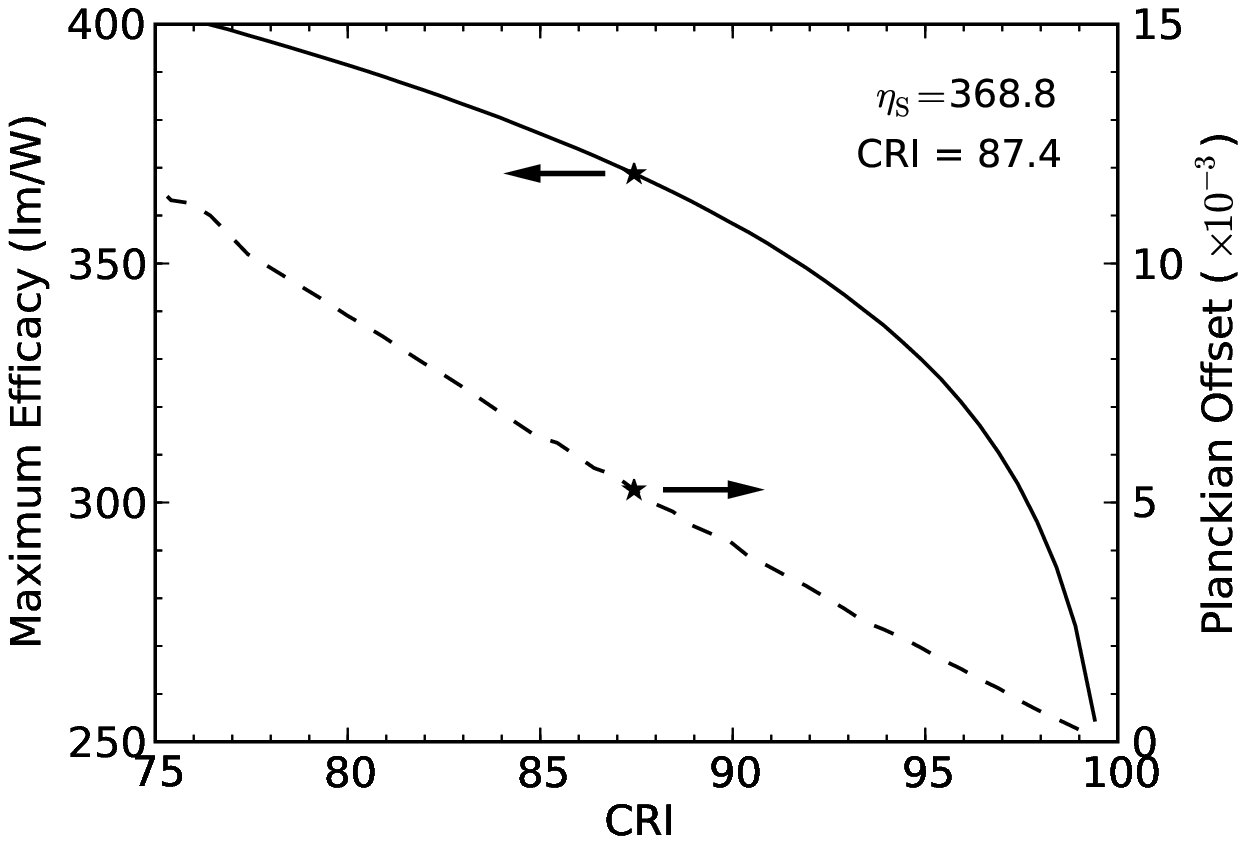}\hfill{}\includegraphics[%
  scale=0.6]{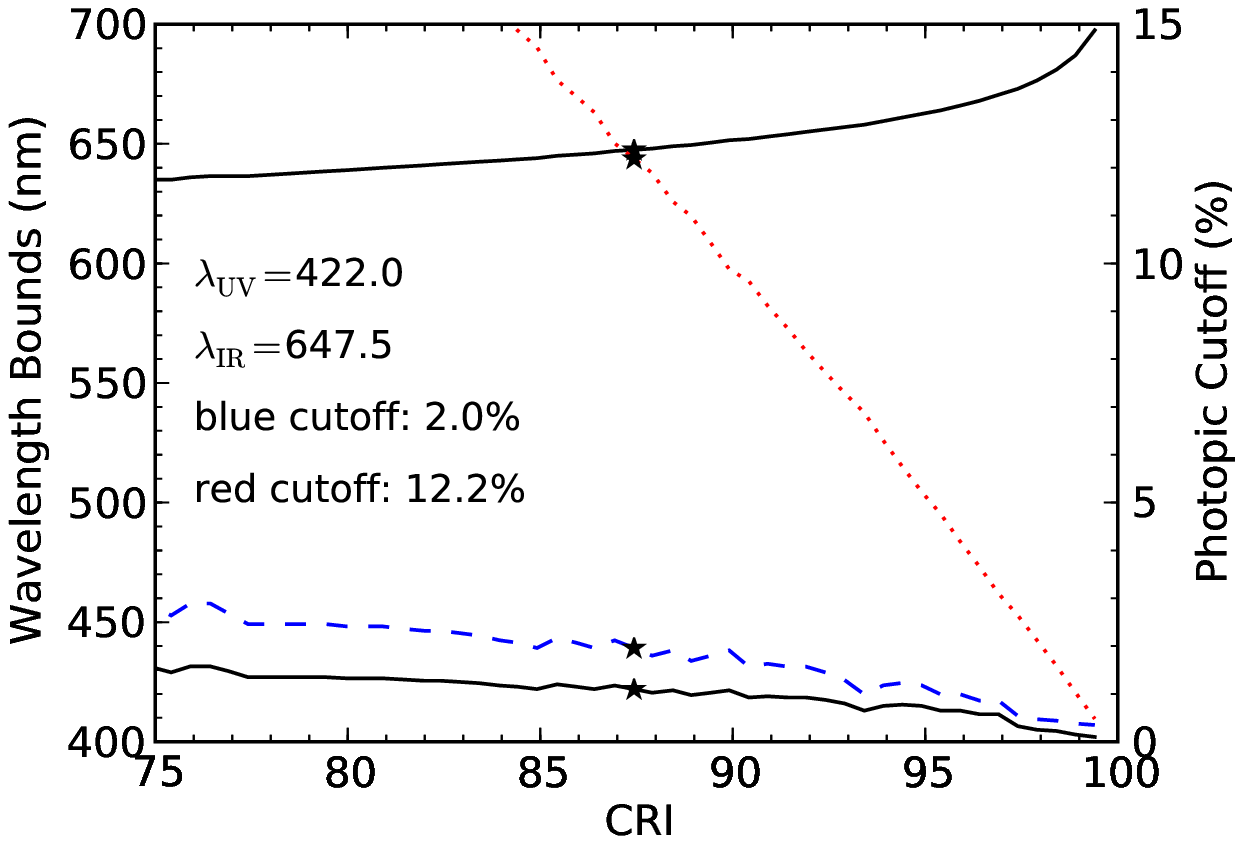}

\caption{(Color online) Conditions for maximum luminous efficacy for a truncated blackbody
at 2800~K, allowing asymmetric red and blue cutoff wavelengths.
Conventions follow those from Fig.~\ref{fig:max58}.
\label{fig:max28}}
\end{figure*}

Relaxing the constraint that $\lambda_{\mathrm{blue}}$ and
$\lambda_{\mathrm{red}}$ be determined by symmetric thresholds on the
photopic sensitivity curves, we can ask what combination of
$\lambda_{\mathrm{blue}}$ and $\lambda_{\mathrm{red}}$ delivers the maximum
luminous efficacy as a function of achieved CRI.  For instance, we can
demand a CRI of 95 and seek the highest $\eta_{\mathrm{S}}$ that can result
from a blackbody function truncated by arbitrary $\lambda_{\mathrm{blue}}$ and
$\lambda_{\mathrm{red}}$.  Fig.~\ref{fig:max58} shows the result for a
truncated 5800~K blackbody.  The Planckian offset exceeds $5.4\times
10^{-3}$ for CRI values below $\sim 94$, at which point the spectral luminous
efficacy is $\eta_{\mathrm{S}}\approx 310$~lm/W.  Exploring what this means
in terms of wavelength cutoff, we find that the maximum luminous efficacy
favors sacrificing red light sooner than giving up blue (right panel of
Fig.~\ref{fig:max58}).  In this case, the
blue cutoff is at 423~nm, corresponding to 2.1\% photopic sensitivity.
Meanwhile, the red cutoff is at 658.5~nm, or a photopic sensitivity of
6.7\%.  Similar trends are found for a truncated 2800~K spectrum, as seen
in Fig.~\ref{fig:max28}---except that the light remains "white" at a lower
CRI, pushing the luminous efficacy up to $\sim 370$~lm/W, sacrificing even
more red light in the red-tilted spectrum.

Exploring one final scenario, one may be tempted to match a light
source to the photopic sensitivity curve, because not much light would
be ``wasted'' at wavelengths to which the human eye is not sensitive.
Such a source can be approximated by a Gaussian curve centered at
560~nm and a full width at half-maximum of 100~nm. This light would
deliver a luminous efficacy of 488~lm/W, but would look distinctly
green, being similar in light distribution to that of green LED, albeit
with a broader spectrum. In particular, the Planckian offset is 0.054,
and the CRI computes to 24: a grossly inadequate source of white light.

\section{Characterization of Real Sources}

We can evaluate today's sources of alternative lighting by acquiring
spectral distributions and computing the corresponding spectral luminous
efficacy.  This technique does not address the efficiency with which
electrical power input is converted to luminous energy,
$\eta_{\mathrm{E}}$, but simply evaluates the efficacy stemming from the
distribution, $\eta_{\mathrm{S}}$.

Spectra were obtained using the USB2000 spectrometer from Ocean Optics,
capturing 12-bit raw data at a spectral resolution between 400--500
and sampling ranging from 0.38~nm at the UV end to 0.28~nm at the
IR end. Spectra are dark-subtracted and calibrated against a solar
spectrum acquired by the same setup that itself is fitted to a 5800~K
Planck function, avoiding absorption features from the solar and terrestrial
atmospheres in the fitting procedure.

Luminous efficacies are evaluated in the range from 400--900~nm,
which does not catch 100\% of the light in all cases (and would be
a poor choice for blackbody sources), but captures the vast majority of the
light output for the sources presented here.

\begin{figure*}
\includegraphics[%
  scale=0.6]{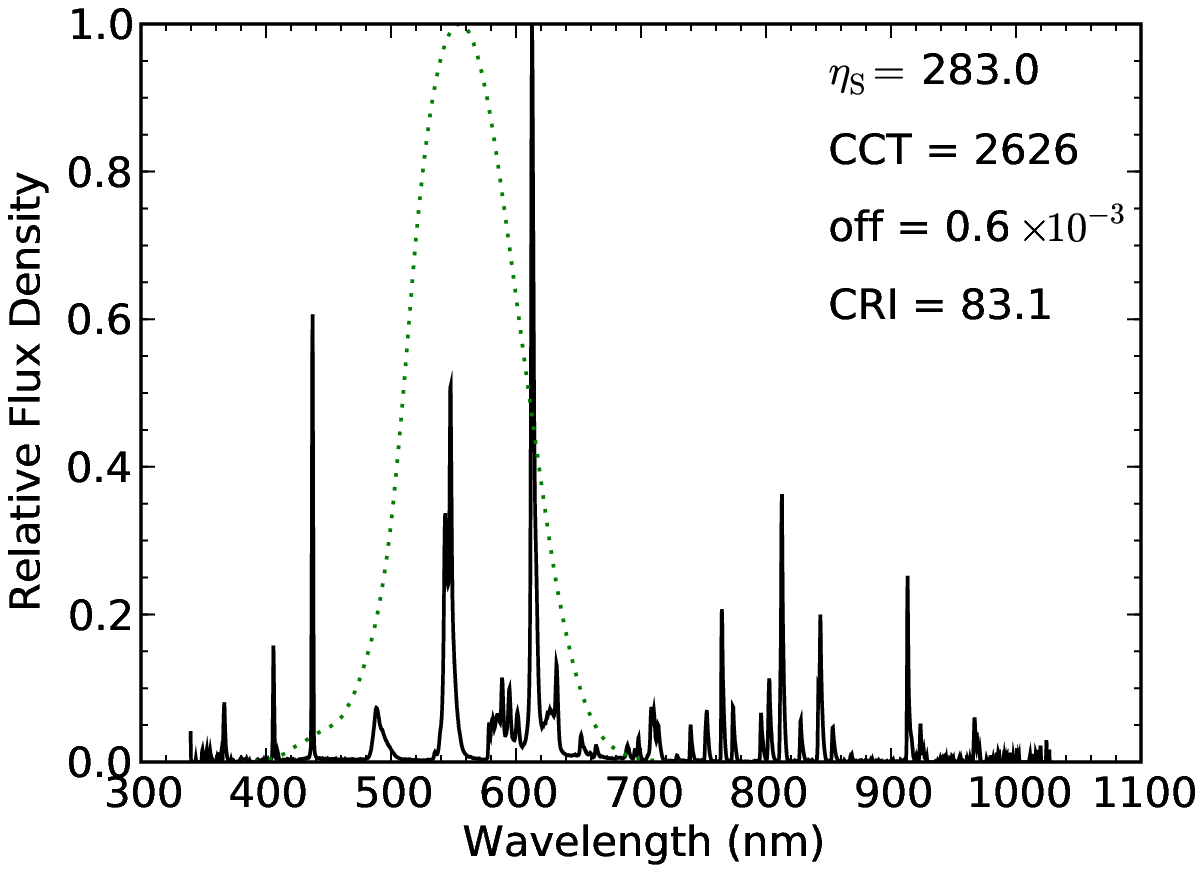}\hfill{}\includegraphics[%
  scale=0.6]{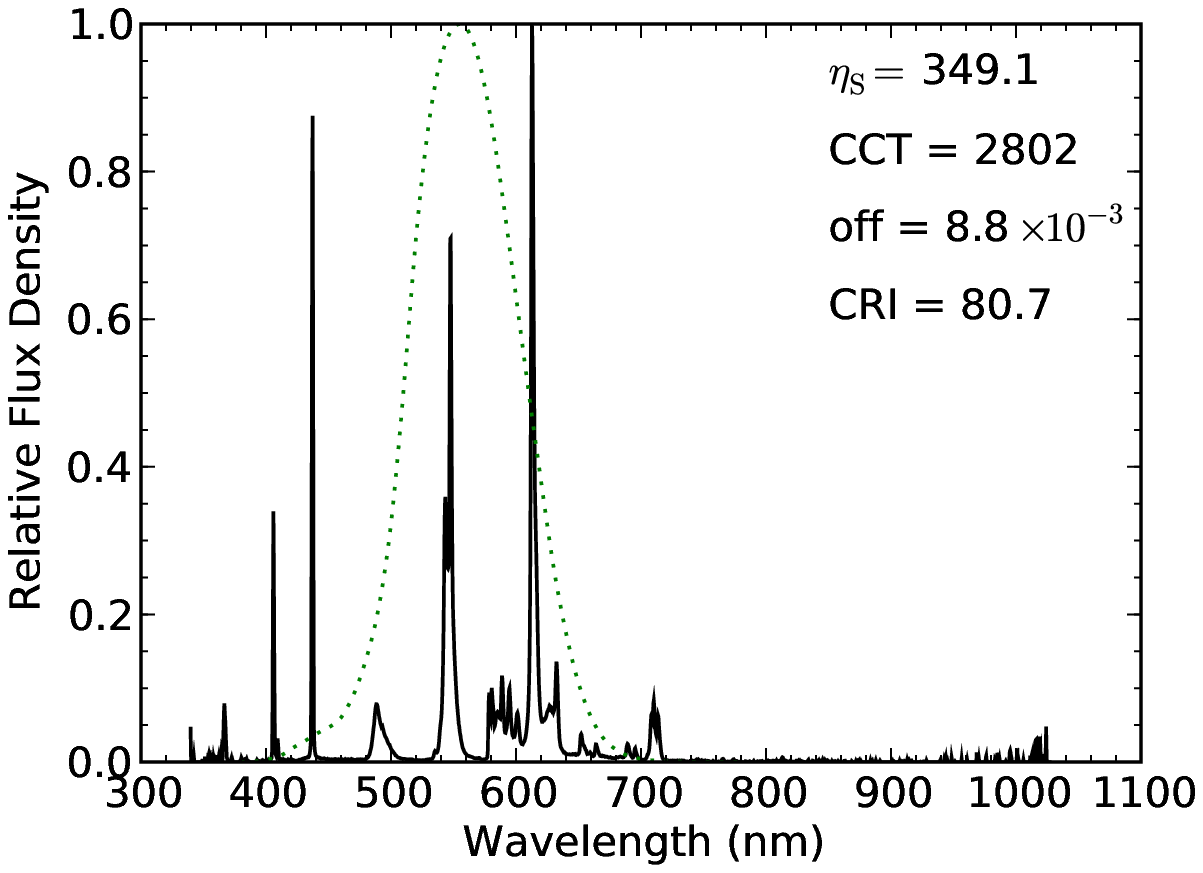}

\caption{16~W CFL (``60~W'' replacement) at turn-on (left) and after
warm-up (right). The spectral luminous efficacy, correlated color
temperature, Planckian offset, and color rendering index are displayed
for each spectrum. The photopic sensitivity curve is also displayed
for reference.\label{fig:CFL-spectra}}
\end{figure*}

Fig.~\ref{fig:CFL-spectra} shows the spectrum of a 16~W compact
fluorescent light rated at 900~lm as it appears seconds after turn-on
from an ambient temperature state and after settling in a warm equilibrium.
The luminous efficacy of the spectral distribution evolves from 283~lm/W
to 349~lm/W as many of the infrared lines disappear and the green
line achieves greater relative dominance.  In the process, the Planckian
offset exceeds the acceptable limit, and the CRI is well short of the
target of 90.  At $\eta_{\mathrm{L}}=56$~lm/W and
$\eta_{\mathrm{S}}=349$~lm/W, the inferred electricity-to-light
ratio is $\eta_{\mathrm{E}}\sim$16\%.

\begin{figure*}
\includegraphics[%
  scale=0.6]{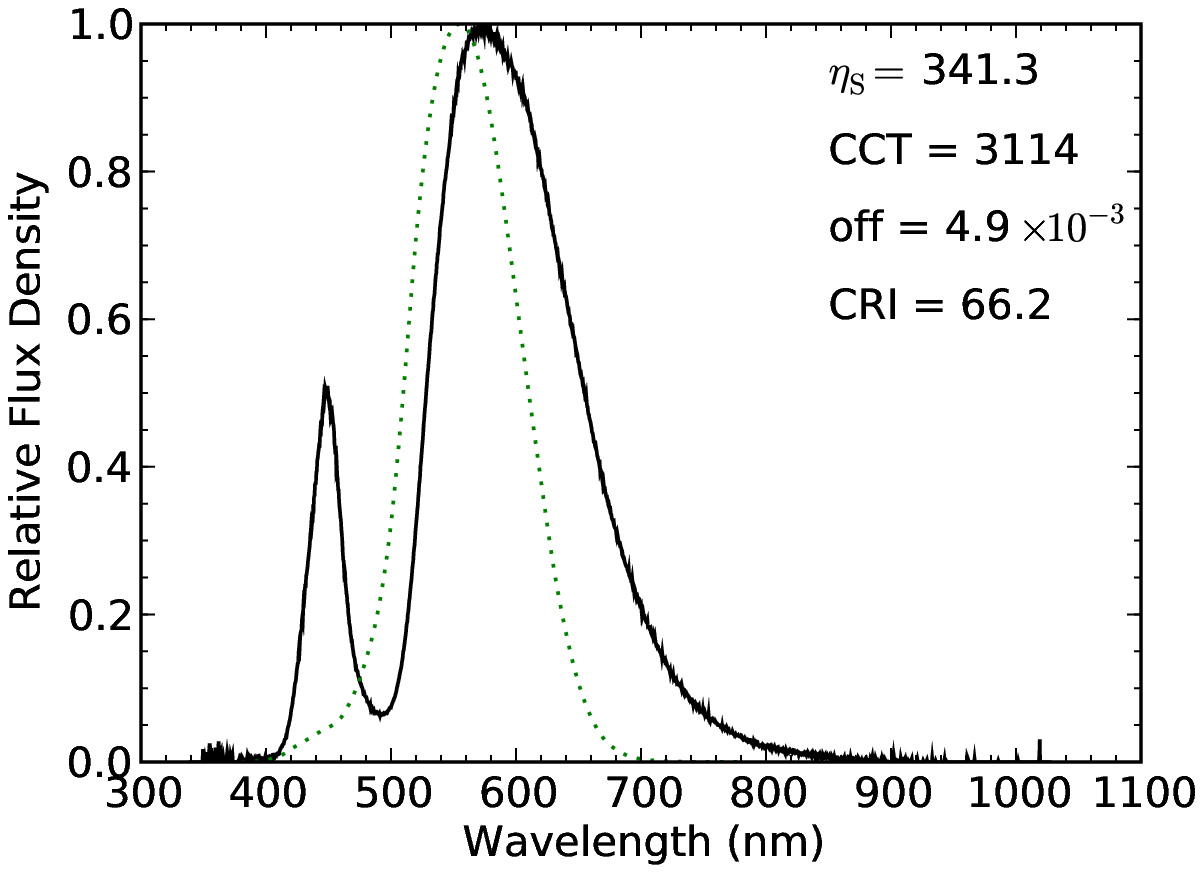}\hfill{}\includegraphics[%
  scale=0.6]{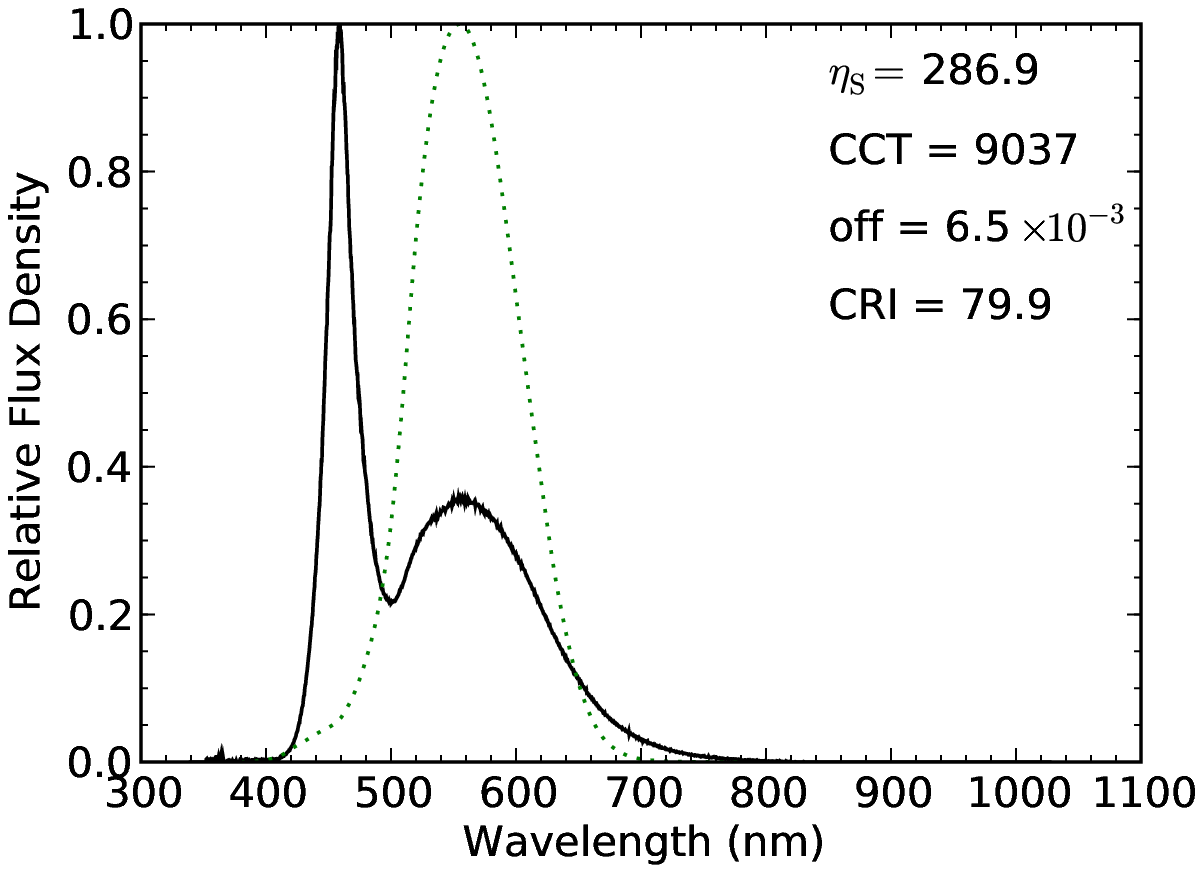}

\caption{1.5~W LED lights labeled as having color temperatures of 3000~K
(left) and 6500~K (right). The measured color temperature of the light at
right is clearly a poor match to that indicated on the
packaging.  Conventions and labels follow that of
Fig.~\ref{fig:CFL-spectra}.  \label{fig:LED-spectra}}

\end{figure*}

Fig.~\ref{fig:LED-spectra} illustrates the spectral distributions of two
lights using similar total electrical power and based on the same blue LED,
but using different phosphors at differing levels of absorption to achieve
advertised color temperatures of 3000~K and 6500~K. The spectral
distributions achieve luminous efficacies of 341~lm/W and R287lm/W,
respectively. While both lights closely bracket the acceptable Planckian
offset, neither have a high CRI---especially the 3000~K light.  The 3000~K
light is rated at 86~lm, implying an electrical-to-light efficiency of
$\eta_{\mathrm{E}}\sim$17\%.

\begin{figure*}
\includegraphics[%
  scale=0.6]{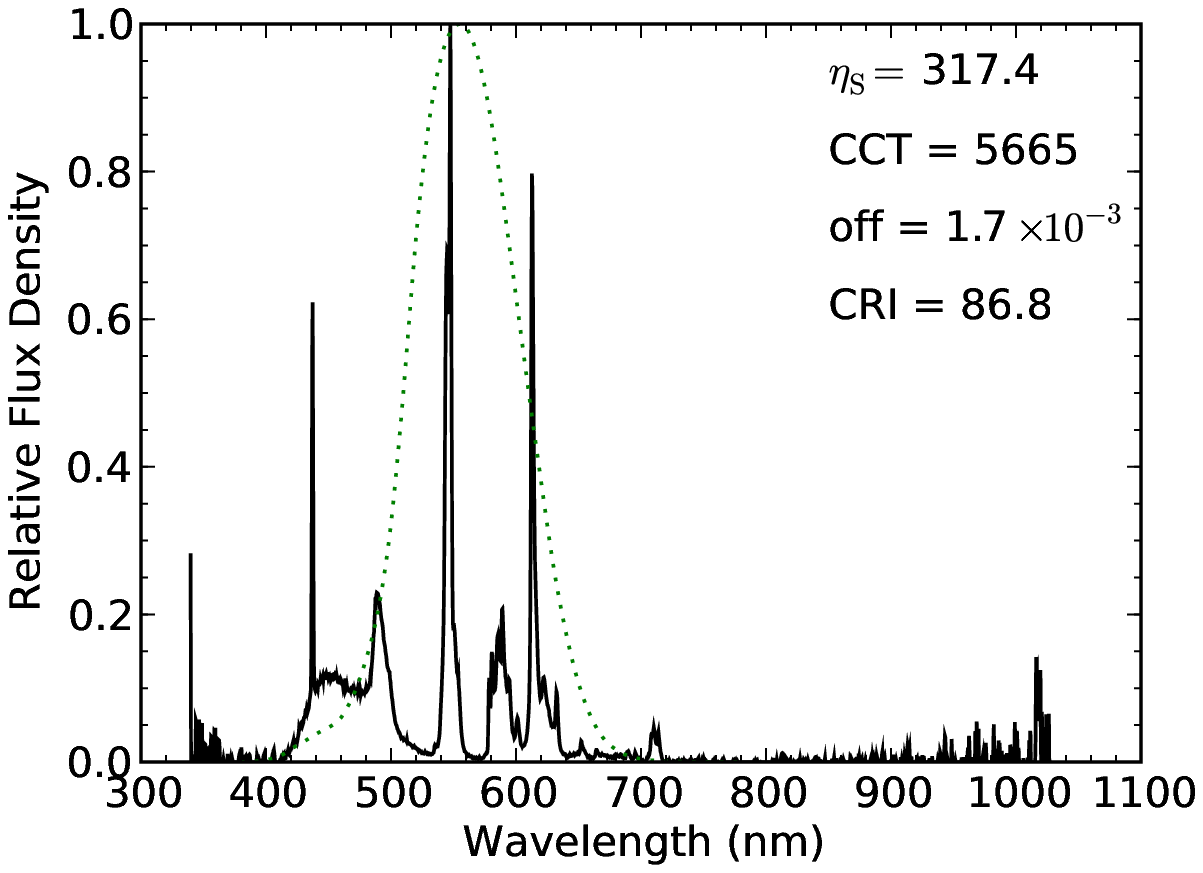}\hfill{}\includegraphics[%
  scale=0.6]{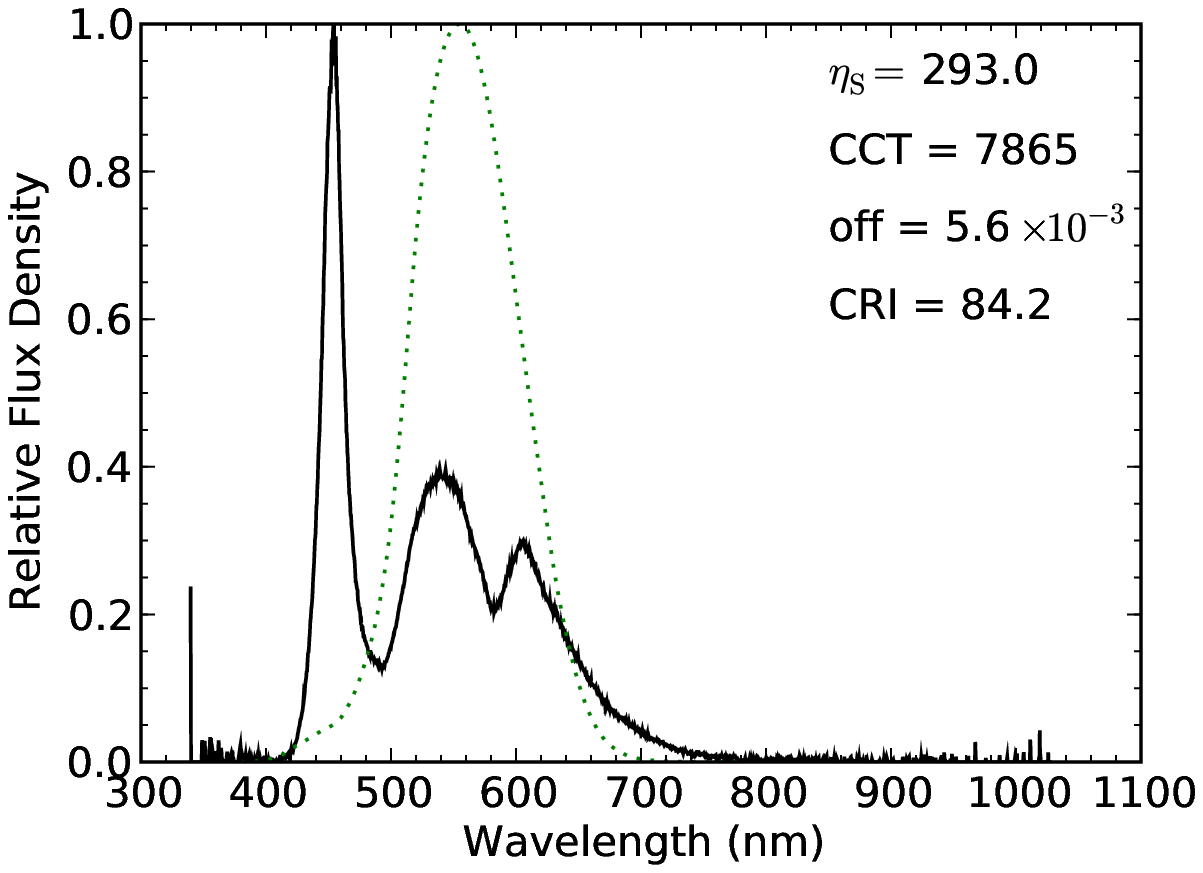}

\caption{Backlighting for laptop computer liquid crystal displays, showing
fluorescent tube illumination (left) and LED illumination (right).
Conventions and labels follow that of Fig.~\ref{fig:CFL-spectra}.\label{fig:LCD-spectra}}
\end{figure*}

Fig.~\ref{fig:LCD-spectra} presents spectra from the white background of
two different laptop computer displays backlit by fluorescent tubes and
LEDs, achieving spectral luminous efficacies of 317~lm/W and 293~lm/W,
respectively. The spectra differ qualitatively from their lighting
counterparts, chiefly in their use of phosphors. Spectra were also acquired
for an LED-illuminated television, scoring 283~lm/W and appearing
qualitatively identical to the spectrum of the LED computer display.  Each
of the displayed spectra have reasonably small Planckian offsets, and the
CRI values are substantially higher than for the lighting counterparts in
Figs.~\ref{fig:CFL-spectra} and \ref{fig:LED-spectra}.

Each of these sources---by confining emission to the visible parts of the
spectrum---are capable of far better spectral luminous efficacies than are
incandescent sources. Each of the spectra presented here land within the
range of 250--350~lm/W, despite vastly different spectral distributions.
For clarity, these lights do not achieve total luminous efficacies above
100~lm/W due to inefficiencies in the \emph{generation} of photons.  Given
that all of the measured spectra in
Figs.~\ref{fig:CFL-spectra}--\ref{fig:LCD-spectra} yield spectral
efficacies in the 280--350~lm/W range, we infer that generation
efficiencies above about $\eta_{\mathrm{E}}\sim 0.3$ in lights sharing
similar spectra will naturally deliver overall luminous efficacies above
100~lm/W.


\section{Conclusions}

Synthesizing a white light source that emits no light outside the visible
band can achieve luminous efficacies in a range between 250--370~lm/W
depending on spectral extent and corresponding ``whiteness."  One
approaches the upper end of this range when truncating the spectrum at the
5\% photopic sensitivity limits of the eye, generating light only between
453--663~nm.  But at this cutoff, the light is already inadequate in terms
of color rendering index and Planckian offset.  By accepting asymmetric
cutoffs, we are able to achieve adequate color properties at
$\eta_{\mathrm{S}}\approx 310$~lm/W for a color temperature of 5800~K, and
$\eta_{\mathrm{S}}\approx 370$~lm/W at 2800~K.  At the high-quality end, it is
unlikely that any white-light application for humans would require
illumination at the very low end of the photopic sensitivity curve, below
0.5\% (corresponding to wavelength cutoffs at 406 and 697~nm), so that
250~lm/W can be taken as the lower bound to maximum practical luminous
efficacy of a white source.

Unlike the pure blackbody, which achieves a maximum luminous efficacy
around 6640~K, a truncated source performs best between 3500--4100~K.
Even so, the spectral efficacy is relatively constant (within 10\%)
across color temperatures ranging from 2500--8000~K for a given 
set of truncation bounds.  Similarly, lower color temperatures produce
higher spectral luminous efficacies under asymmetric cutoff wavelengths.

Leaving aside the efficiency with which photons are generated, any
real white-like spectrum confining itself to visible wavelengths is likely
to achieve a \emph{spectral} luminous efficacy in the range of 250--350~lm/W,
as is demonstrated by a wide variety of sources above. This fact emphasizes
that improvements in the efficacy of current lighting technology must
primarily involve advances in photon generation efficiency rather than
spectral conditioning.

\end{document}